\documentclass[manuscript]{aastex}
\usepackage{amsmath,amssymb}
\usepackage{epstopdf}
\usepackage{graphicx}
\usepackage{xcolor}

\usepackage[titletoc,title]{appendix}
\title{Turbulence Transport Modeling and First Orbit Parker Solar Probe (PSP) Observations}

\newcommand{\myemail}

\usepackage[inline]{enumitem}
\usepackage{multirow}
\usepackage{lscape}

\shorttitle{}
\shortauthors{Adhikari et al.}

\begin{document}

\begin{abstract}
Parker Solar Probe (PSP) achieved its first orbit perihelion on November 6, 2018, reaching a heliocentric distance of about 0.165 au (35.55 R$_\odot$). Here, we study the evolution of fully developed turbulence associated with the slow solar wind along the PSP trajectory between 35.55 R$_\odot$ and 131.64 R$_\odot$ in the outbound direction, comparing observations to a theoretical turbulence transport model. Several turbulent quantities, such as the fluctuating kinetic energy and the corresponding correlation length, the variance of density fluctuations, and the solar wind proton temperature are determined from the PSP SWEAP plasma data along its trajectory between 35.55 R$_\odot$ and 131.64 R$_\odot$. The evolution of the PSP derived turbulent quantities are compared to the numerical solutions of the nearly incompressible magnetohydrodynamic (NI MHD) turbulence transport model recently developed by Zank et al. (2017). We find reasonable agreement between the theoretical and observed results. On the basis of these comparisons, we derive other theoretical turbulent quantities, such as the energy in forward and backward propagating modes, the total turbulent energy, the normalized residual energy and cross-helicity, the fluctuating magnetic energy, and the correlation lengths corresponding to forward and backward propagating modes, the residual energy, and the fluctuating magnetic energy.
\end{abstract}
	
	\author{L. Adhikari$^{1}$, G. P. Zank$^{1,2}$, L.-L. Zhao$^{1}$, J. C. Kasper$^{3,4}$, K. E. Korreck$^4$, M. Stevens$^4$, A.W Case$^4$, P. Whittlesey$^5$, D. Larson$^5$, R. Livi$^5$, and K. G. Klein$^{6,7}$}
	
	\altaffiltext{1}{Center for Space Plasma and Aeronomic Research (CSPAR), University of Alabama in Huntsville, Huntsville, AL 35899, USA }
	
	\altaffiltext{2}{Department of Space Science, University of Alabama in Huntsville, Huntsville, AL 35899, USA}	
	
	\altaffiltext{3}{Climate and Space Sciences and Engineering, University of Michigan, Ann Arbor, MI 48109, USA}
	\altaffiltext{4}{Smithsonian Astrophysical Observatory, Cambridge, MA 02138, USA}	
	\altaffiltext{5}{Space Sciences Laboratory, University of California, Berkeley, CA 94720-7450, USA}
	\altaffiltext{6}{Lunar and Planetary Laboratory, University of Arizona, Tucson, AZ 85721, USA}
	\altaffiltext{7}{Department of Planetary Sciences, University of Arizona, Tucson, AZ 85719, USA}

\section{Introduction}
Parker Solar Probe (PSP) \citep{2016SSRv..204....7F} has been exploring the inner heliosphere since its launch on August 12, 2018. To date, PSP has executed two of its twenty four perihelion passes close to the Sun, enabling a comparison of inner heliopsheric observations with theoretical predictions. PSP achieved its first orbit perihelion on November 6, 2018, reaching a heliocentric distance of about 0.165 au (35.55 R$_\odot$). The inbound and outbound direction of the PSP trajectory at its closest point of approach is almost radial with respect to the solar wind flow. Since the PSP trajectory approaches the Sun more and more closely with each orbit, PSP provided data from each orbit enables us to study the evolution of turbulence in the inner heliosphere. 

In this manuscript, we study the evolution of turbulence between the perihelion of the first orbit of the PSP (0.165 au or 35.55 R$_\odot$) and 131.64 R$_\odot$ in the outbound direction using the nearly incompressible turbulence transport model equations of \cite{2017ApJ...835..147Z} and the PSP SWEAP \citep{2016SSRv..204..131K} plasma measurements. \cite{2017ApJ...835..147Z} developed coupled turbulence transport model equations for the majority quasi-2D and minority slab turbulence, which have been successfully compared with \textit{Voyager} 2, \textit{Ulysses}, and \textit{New Horizons Solar Wind Around Pluto (NH SWAP)} instruments in the outer heliosphere beyond 1 au, and with \textit{Helios} 2 in the inner heliosphere within 1 au \citep{2017ApJ...835..147Z,2018ApJ...869...23Z,2015ApJ...805...63A,2017ApJ...841...85A}. The nearly incompressible turbulence transport model equations are also employed to study turbulence in the solar corona \citep{2018ApJ...854...32Z,2019ApJ...876...26A,adhikari2019}. The nearly incompressible Zank et al. (2017) model can predict several quasi-2D and slab turbulent quantities that include both the fluctuating velocity and magnetic field (see Table 1).
\begin{table}[h!]
	\centering
	\begin{tabular}{|c |c|}
		\hline
		\multicolumn{2}{|c|}{ \textbf{Turbulent quantities}} \\		\hline
		$\langle {z^\pm}^2 \rangle$, $\langle {{z^\infty}^\pm}^2 \rangle$, $\langle {{z^*}^\pm}^2 \rangle$ &   \shortstack{Energy in forward and backward \\ propagating modes total, quasi-2D, slab}  \\ \hline
		$E_T$, $E_T^\infty$, $E_T^*$ &  \shortstack{Turbulent energy total, quasi-2D, slab}  \\ \hline
		$E_D/\sigma_D$, $E_D^\infty/\sigma_D^\infty$, $E_D^*/\sigma_D^*$ & \quad \shortstack{Residual/normalized residual energy \\ total, quasi-2D,  slab }  \\ \hline
		$E_C/\sigma_c$, $E_C^\infty/\sigma_c^\infty$, $E_C^*/\sigma_c^*$ &  \shortstack{ Cross-helicity/normalized cross-helicity \\ total, quasi-2D, slab}  \\ \hline
		$\langle B^2 \rangle$, $\langle {B^\infty}^2 \rangle$, $\langle {B^*}^2 \rangle$ &  \shortstack{ Fluctuating magnetic energy \\ total, quasi-2D, slab} \\ \hline
		$r_A$, $r_A^\infty$, $r_A^*$ & Alfv\'en ratio total, quasi-2D, slab \\ \hline
		$\langle u^2 \rangle$, $\langle {u^\infty}^2 \rangle$, $\langle {u^*}^2 \rangle$ &  \shortstack{ Fluctuating kinetic energy \\ total, quasi-2D, slab} \\ \hline
		$L_{\infty,*}^\pm$, $L_D^{\infty,*}$ & \shortstack{Quasi-2D, slab (for forward and backward propagating modes), \\ and residual energy correlation function}  \\ \hline
		$\lambda_{\infty,*}^\pm$, $\lambda_D^{\infty,*}$ & \shortstack{Quasi-2D and slab correlation length for \\
			forward and backward propagating modes, \\ residual energy } \\ \hline
		$l_b^\infty$, $l_b^*$ & \shortstack{Correlation length of magnetic \\ field fluctuations quasi-2D, slab} \\ 
		\hline
		$l_u^\infty$, $l_u^*$ & \shortstack{Correlation length of velocity \\ fluctuations quasi-2D, slab} \\ 
		\hline
		$\langle {\rho^\infty}^2 \rangle$ & \shortstack{Variance of density fluctuations} \\ 
		\hline	
		$T$ & \shortstack{Solar wind proton temperature} \\ 
		\hline	
	\end{tabular}
	\caption{Turbulent velocity and magnetic field quantities predicted by Zank et al. (2017). }
\end{table}
However, in this manuscript we only compare the plasma quantities predicted by the model with the PSP SWEAP plasma measurements  \citep{2016SSRv..204..131K}, such as the fluctuating kinetic energy and the correlation length of velocity fluctuations, the variance of density fluctuations, and the solar wind proton temperature. We then derive the theoretical solutions of the energy in forward and backward propagating modes, the fluctuating magnetic energy, the normalized cross-helicity and residual energy, the Alfv\'en ratio, the correlation lengths of forward and backward propagating modes and residual energy, and the correlation length of magnetic field fluctuations between 35.55 R$_\odot$ and 131.64 R$_\odot$. These predictions will be compared to PSP observations when the magnetic field data become available.

Our understanding of solar wind turbulence improves with the availability of measurements from several spacecraft. The solar wind has been used to study magnetohydrodynamic (MHD) turbulence since the late 1960s \citep{1968ApJ...153..371C,1971JGR....76.3534B, 1982JGR....87.3617B, 1995ESASP.371..137G,1995ARA&A..33..283G,1995SSRv...73....1T,2005LRSP....2....4B,2013LRSP...10....2B}. \cite{1968ApJ...153..371C} and \cite{1971JGR....76.3534B} studied the MHD fluctuation in the solar wind, and found that there exists MHD wave \citep{1971JGR....76.3534B} and MHD turbulence \citep{1968ApJ...153..371C}. The velocity near the Sun is usually highly correlated with fluctuations in magnetic fields \citep{1967P&SS...15..953C,1971JGR....76.3534B}, and possesses a high degree of Alfv\'enicity, which decreases with increasing heliocentric distance  \citep{1987JGR....9212023R,1987JGR....9211021R}. 

As the PSP approaches closer to the Sun with each orbit, we have the opportunity to study the properties of fluctuations of velocity and magnetic field and their dissipation mechanism in the deeper inner heliosphere. PSP will promote our understanding of the turbulence in the inner heliopshere, and improve our understanding of coronal heating and the origin of solar wind in open magnetic field regions \citep{1999ApJ...523L..93M,2001ApJ...548..482D,2002ApJ...575..571D,2001ApJ...551..565O,2005ApJ...632L..49S,2007ApJS..171..520C,2009ApJ...707.1659C,2010ApJ...720..503C,2013ApJ...767..125C,2010ApJ...708L.116V,2000ApJ...528..509V,2014ApJ...787..160W,2018ApJ...854...32Z,2010ApJ...720..824C,2011ApJ...736....3V,2016ApJ...821..106V}. In the outer heliosphere, the solar wind proton temperature is non-adiabatic, which is believed to be caused by the dissipation of turbulence \citep{1994JGR....99.6561G,1988GeoRL..15...88F,1995JGR...10017059W,1999PhRvL..82.3444M,2001JGR...106.8253S,2006ApJ...638..508S,2006JGRA..111.9111S,2003ApJ...592..564I,2005ApJ...623..502I,2008JGRA..113.8105B,2010ApJ...719..716I,2010JGRA..115.2101N,2011ApJ...727...84U,2011JGRA..116.8105O,2014ApJ...793...52A,2015ApJ...805...63A,2017ApJ...841...85A,2017ApJ...835..147Z,2018ApJ...869...23Z,2017ApJ...837...75S,2015ApJ...805..155W,2016ApJ...833...17W}.

We organize the manuscript as follows. In Section 2, we discuss the quasi-2D and slab turbulence transport model equations. Section 3 presents the comparison between the theoretical results and observed results derived from PSP SWEAP plasma measurements. Section 4 discusses the other theoretical solutions of the turbulence transport model equations that includes the magnetic field. Finally, Section 5 presents conclusions from this manuscript.

\section{Theoretical model equations}
The fluctuating solar wind speed and magnetic field observed by PSP during its first two perihelia possess many features of fully developed MHD turbulence. The small scale fluctuating fields are most naturally described by the Els\"asser variables, $\textbf{z}^\pm = \textbf{u}\pm \textbf{B}/ {\sqrt{\mu_0 \rho}}$ \citep{PhysRev.79.183}, where $\rho$ is the solar wind density, ${\bf u}$ is the fluctuating velocity, ${\bf B}$ is the fluctuating magnetic field, and $\mu_0$ is the magnetic permeability. The Els\"asser variables $\bf z^\pm$ are functions of both large-scales (e.g., background solar wind scales) and small-scales (e.g., turbulence scales) \citep[see,][]{1990JGR....9514863Z,1990JGR....9514881Z,1996JGR...10117093Z,2012ApJ...745...35Z,2017ApJ...835..147Z}, and are important parameters for describing MHD turbulence.

Observations of evident turbulent behavior in the solar wind (e.g., Kolmogorov-like spectra in energy, magnetic field, temperature and density fluctuations or the non-adiabatic heating of the solar wind, for example) are interpreted in terms of an incompressible MHD turbulence phenomenology. The convergence of compressible MHD to an incompressible state is achieved via a singular perturbation expansion based on the existence of a small turbulent sonic Mach number $M_s=\delta u/C_s$, where $\delta u$ is the characteristic fluctuating plasma speed and $C_s$ the local sound speed. For a homogeneous system, NI MHD theory predicts that the density fluctuations scales as $\delta \rho \sim O(M_s^2)$ \citep{1991PhFl....3...69Z,1992JGR....9717189Z}, whereas for an inhomogeneous system $\delta \rho \sim O(M_s)$ \citep{2010ApJ...718..148H,1998ApJ...494..409B}. In the inhomogeneous solar wind, we might expect the latter scaling to hold \citep{1994JGR....9921481T,1995JGR...100.9475B,1995JGR...100.5871B,1993JGR....98.7837K}. We evaluate the scaling of density fluctuations observed by PSP using SWEAP data during slow solar wind encounters, finding that indeed $\delta \rho \sim O(M_s)$ (see Appendix A). We find too that the $\delta \rho/\rho$, where $\rho$ is the mean local density, is typically in the range 0.1--0.2 during the PSP encounter of the slow solar wind. These results suggest that indeed the slow solar wind observed by PSP during its first encounter was in a nearly incompressible state.

If we adopt a nearly incompressible description of low-frequency MHD turbulence in the solar wind \citep{1992JGR....9717189Z,1993PhFl....5..257Z,2010ApJ...718..148H,2012ApJ...745...35Z,2017ApJ...835..147Z,2014ApJ...793...52A,2015ApJ...805...63A,2017ApJ...851..117A}, the plasma beta $\beta_p \gg 1$ regime ($\beta_p = P/(B^2/{2 \mu_0})$, where $P$ is the thermal plasma pressure and $B$ is the total magnetic field) corresponds to isotropic turbulence even in the presence of a mean magnetic field \citep{1992JGR....9717189Z,1993PhFl....5..257Z,1996JGR...10117093Z,2012ApJ...745...35Z}, whereas the $\beta_p \sim 1$ and $\ll 1$ regimes show that low-frequency MHD turbulence can be decomposed into a majority quasi-2D component and a minority slab component. 
As we describe below, away from the heliospheric current sheet, the region observed by PSP that we consider from the perspective of turbulence transport modeling possess $\beta_p$ values that are typically $\sim 1$ and $\ll 1$. In a related paper, Zhao et al. (2019) (this issue) identify numerous magnetic flux ropes of various size, including small-scales in the PSP slow wind data considered here, providing evidence for the idea that fully developed turbulence in the slow wind possesses probably a majority quasi-2D component. The fast wind that Zhao et al. (2019) investigate is dominated largely by outwardly propagating Alfv\'en waves, as is expected from certain models of turbulence in the solar corona \citep{2018ApJ...854...32Z,2019ApJ...876...26A}. For a $\beta_p \sim 1$ or $\ll 1$ plasma, the total Els\"asser variables can be further decomposed as the sum of quasi-2D and slab Els\"asser variables, i.e., ${\bf z}^ \pm = {{\bf z}^\infty}^\pm + {{\bf z}^*}^\pm$ provided certain symmetries of the underlying turbulence are present \citep{2017ApJ...835..147Z}. The difference between ${{\bf z}^\infty}^\pm$ and ${{\bf z}^*}^\pm$ reflects the anisotropy of the solar wind fluctuations in the energy-containing range. However, in the 2D plane perpendicular to the magnetic field, the quasi-2D component is isotropic, while the slab component is axisymmetric with a particular direction $\hat{r}$ defined by the magnetic field. In this case, the net geometry of the NI MHD system is a superposition of quasi-2D (isotropic) and slab (axisymmetric) turbulence.
The majority quasi-2D and minority slab Els\"asser variables can be written as \citep{2017ApJ...835..147Z}
\begin{equation}
{{\bf z}^\infty}^\pm = {\bf u}^\infty \pm \frac{{\bf B}^\infty}{\sqrt{\mu_0 \rho}}~ \text{and} ~ {{\bf z}^*}^\pm = {\bf u}^* \pm \frac{{\bf B}^*}{\sqrt{\mu_0 \rho}},
\end{equation} 
where the superscript \textquotedblleft$\infty$\textquotedblright~refers to quasi-2D turbulence, and the superscript \textquotedblleft*\textquotedblright denotes slab turbulence. The quasi-2D and slab variances of the Els\"asser variables, and the residual energy $E_D$ can be written as \citep{2012ApJ...745...35Z,2017ApJ...835..147Z},
\begin{equation}
\langle { {z^{\infty,*} }^\pm }^2 \rangle  = \langle { {{\bf z}^{\infty,*} }^\pm } \cdot { {{\bf z}^{\infty,*} }^\pm } \rangle;~E_D^{\infty,*} = \langle { {{\bf z}^{\infty,*} }^+ } \cdot { {{\bf z}^{\infty,*} }^- } \rangle,
\end{equation}  
where $\langle { {z^{\infty,*} }^+ }^2 \rangle$ and $\langle { {z^{\infty,*} }^- }^2 \rangle$ denote the energy in quasi-2D/slab forward and backward propagating modes, respectively. Similarly, the correlation functions corresponding to forward/backward propagating modes, and the residual energy can be written as
\begin{equation}
\begin{split}
& L_{\infty,*}^\pm = \int \langle {{\bf z}^{\infty,*}}^\pm \cdot {{{\bf z}^{\infty,*}}^\pm}' \rangle dy \equiv \langle {{z^{\infty,*}}^\pm}^2 \rangle \lambda^\pm_{\infty,*};\\
&L_D^{\infty,*} = \int  \langle { {{\bf z}^{\infty,*} }^+ } \cdot { {{\bf z}^{\infty,*} }^- }' + { {{\bf z}^{\infty,*} }^+ }' \cdot { {{\bf z}^{\infty,*} }^- }  \rangle dy \equiv E_D^{\infty,*} \lambda_D^{\infty,*},
\end{split}
\end{equation}   
where $y=|{\bf y}|$ is the spatial lag between fluctuations, and ${ {{\bf z}^{\infty,*} }^- }'$ denotes the lagged Els\"asser variables. The parameters $\lambda_{\infty,*}^\pm$ and $\lambda_D^{\infty,*}$ are the quasi-2D/slab correlation lengths corresponding to forward/backward propagating modes, and the residual energy. Considering the conservation of magnetic flux $r^2 B = \text{const} = r_0^2 B_0$, the magnetic field can be expressed as
\begin{equation*}
\textbf{B} = B_0 \bigg(\frac{r_0}{r} \bigg)^2 \hat{r},
\end{equation*}
where $B_0$ is the magnetic field at the reference point $r_0$, and $\hat{r}$ is the direction of the magnetic field.  

\cite{2017ApJ...835..147Z} developed transport equations for the evolution of the majority quasi-2D and minority slab turbulence on the basis of a nearly incompressible phenomenology \citep{2010ApJ...718..148H}. 
The 1D steady-state majority quasi-2D turbulence transport model equations in a spherically symmetrical coordinate $r$ can be written as \citep{2017ApJ...835..147Z,2018ApJ...854...32Z,2017ApJ...841...85A},
\begin{equation}
\begin{split}
U \frac{d \langle {{z^\infty}^\pm}^2 \rangle}{d r} +  \frac{U}{r}  \bigg(\langle {{z^\infty}^\pm}^2 \rangle + E_D^\infty \bigg)   =- 2 \alpha \frac{ \langle {{z^\infty}^\pm}^2 \rangle^2  \langle {{z^\infty}^\mp}^2 \rangle^{1/2}}{L_\infty^\pm} \\
+ 2 \alpha\frac{ \langle {{z^*}^\pm}^2 \rangle^2  \langle {{z^*}^\mp}^2 \rangle^{1/2}}{L_*^\pm} + 2{C^\pm_{sh}} \frac{r_0 |\Delta U| V_{A0}^2}{r^2};
\end{split}
\end{equation}
\begin{equation}
\begin{split}
U \frac{d E_D^\infty}{d r} + \frac{U}{r} \big(E_D^\infty + E_T^\infty \big)   =-\alpha E_D^\infty \bigg( \frac{ \langle {{z^\infty}^+}^2 \rangle  \langle {{z^\infty}^-}^2 \rangle^{1/2}}{L_\infty^+} + \frac{ \langle {{z^\infty}^-}^2 \rangle  \langle {{z^\infty}^+}^2 \rangle^{1/2}}{L_\infty^-}  \bigg) \\
+ \alpha E_D^* \bigg( \frac{ \langle {{z^*}^+}^2 \rangle  \langle {{z^*}^-}^2 \rangle^{1/2}}{L_*^+} + \frac{ \langle {{z^*}^-}^2 \rangle  \langle {{z^*}^+}^2 \rangle^{1/2}}{L_*^-}  \bigg) + 2{C^{E_D}_{sh}} \frac{r_0 |\Delta U| V_{A0}^2}{r^2};
\end{split}
\end{equation}
\begin{equation}
U \frac{d L_\infty^\pm}{d r} + \frac{U}{r} \big(L_\infty^\pm + \frac{L_D^\infty}{2} \big)  =0;
\end{equation}
\begin{equation}
U\frac{dL_D^\infty}{dr} + \frac{2U}{r}  \bigg(L_D^\infty + L_\infty^+ + L_\infty^- \bigg) = 0,
\end{equation}
where $E_T^\infty = (\langle {{z^\infty}^+}^2 \rangle + \langle {{z^\infty}^-}^2 \rangle)/2$ is the total turbulent energy of the quasi-2D fluctuations, and $\alpha$ is the von K\'arm\'an-Taylor constant. \cite{2007JGRA..112.7101V, 2018ApJ...853..153M} suggest that $\alpha$ is about 1/10 for turbulence in the heliosphere, whereas \cite{2016ApJ...821..106V} suggest that a constant $\alpha$ might not be a good approximation in the sub-Alfv\'enic solar wind. Different authors have chosen different values for $\alpha$. For example, \cite{1999ApJ...523L..93M} chose $\alpha=1$ in their model of coronal heating by magnetohydrodynamic turbulence (see also \cite{2018ApJ...854...32Z}). In our study, we consider $\alpha=0.1$. The parameter $\alpha$ controls the cascade rate of turbulence and can therefore result in different radial profiles of the turbulence energy because the intensity of the nonlinear terms in the turbulence model is governed by whether $\alpha$ is large or small. The parameters $C^\pm_{sh}$ and $C_{sh}^{E_D}$ are parametrized strengths of the shear source of energy in forward and backward modes and the residual energy ($E_D$), respectively. In Equations (4) and (5), the third term on the right hand side is the shear driving source of turbulence. Similarly, the 1D steady-state turbulence transport equations for the minority slab turbulence are \citep{2017ApJ...835..147Z,2018ApJ...854...32Z,2017ApJ...841...85A},
\begin{equation}
\begin{split}
\bigg(U \mp V_{A0} \frac{r_0}{r} \bigg) \frac{d \langle {{z^*}^\pm}^2 \rangle}{d r}  - (2b-1)\frac{U}{r} \langle {{z^*}^\pm}^2 \rangle + (6b -1)\frac{U}{r} E_D^* 
\pm (4b-1) \frac{V_{A0}}{r_0} \bigg(\frac{r}{r_0} \bigg)^2 E_D^* \\
 \pm \frac{V_{A0}}{r_0} \bigg( \frac{r_0}{r} \bigg)^2 \langle {{z^*}^\pm}^2 \rangle  =-2\alpha\frac{ \langle {{z^*}^\pm}^2 \rangle \langle {{z^\infty}^\pm}^2 \rangle  \langle {{z^\infty}^\mp}^2 \rangle^{1/2}}{L_\infty^\pm} 
- 2\alpha\frac{ \langle {{z^*}^\pm}^2 \rangle^2  \langle {{z^*}^\mp}^2 \rangle^{1/2}}{L^\pm_*}  + 2{{C^*}^\pm_{sh}} \frac{r_0 |\Delta U| V_{A0}^2}{r^2};
\end{split}
\end{equation}
\begin{equation}
\begin{split}
& U \frac{dE_D^*}{dr} - (2b -1) \frac{U}{r} E_D^* + (6b -1) \frac{U}{r} E_T^*  - (4b-1) E_C^* \frac{V_{A0}}{r_0} \bigg(\frac{r_0}{r} \bigg)^2 
=  - \alpha E_D^* \bigg( \frac{ \langle {{z^\infty}^+}^2 \rangle  \langle {{z^\infty}^-}^2 \rangle^{1/2}}{L_\infty^+} \\
& + \frac{ \langle {{z^\infty}^-}^2 \rangle  \langle {{z^\infty}^+}^2 \rangle^{1/2}}{L_\infty^-}  \bigg)  - \alpha E_D^* \bigg( \frac{ \langle {{z^*}^+}^2 \rangle  \langle {{z^*}^-}^2 \rangle^{1/2}}{L^+_*} + \frac{ \langle {{z^*}^-}^2 \rangle  \langle {{z^*}^+}^2 \rangle^{1/2}}{L^-_*}  \bigg)+ 2{{C^*}^{E_D}_{sh}} \frac{r_0 |\Delta U| V_{A0}^2}{r^2};
\end{split}
\end{equation}
\begin{equation}
\begin{split}
\bigg(U \mp {V_{A0}} \frac{r_0}{r} \bigg) \frac{d L_*^\pm}{d r} -(2b-1)   \frac{U}{r} L_*^\pm + \bigg({3}b - \frac{1}{2}\bigg) \frac{U}{r} L_D^* \pm \bigg(2b-\frac{1}{2}\bigg)\frac{V_{A0}}{r_0} \bigg(\frac{r_0}{r}\bigg)^2 L_D^* \\
\pm \frac{V_{A0}}{r_0}\bigg(\frac{r_0}{r}\bigg)^2 L_*^\pm  =0;
\end{split}
\end{equation}
\begin{equation}
\begin{split}
U \frac{d L_D^*}{d r} - (2b-1) \frac{U}{r} L_D^*+ 2\bigg(3b-\frac{1}{2}\bigg) \frac{U}{r} (L_*^+ + L_*^-) - 2\bigg(2b-\frac{1}{2} \bigg)(L_*^+ - L_*^-) \frac{V_{A0}}{r_0}\bigg(\frac{r_0}{r}\bigg)^2 = 0,
\end{split}
\end{equation}
where $V_{A0}$ is the Alfv\'en velocity at a reference $r_0$. We use $b=0.26$ (see Zank et al. (2012) for further discussion of this value). The quantity $E_T^* = (\langle {{z^*}^+}^2 \rangle + \langle {{z^*}^-}^2 \rangle)/2$ is the total turbulent energy of the slab fluctuations, and $E_C^* = (\langle {{z^*}^+}^2 \rangle - \langle {{z^*}^-}^2 \rangle)/2$, the energy difference between that in forward and backward propagating modes, is the cross-helicity of slab turbulence. The parameters ${C^*}^\pm_{sh}$ and ${C^*}^{E_D}_{sh}$ are the strengths for a stream-shear source of slab turbulence, and the third term of the right hand side of Equations (8) and (9) is the shear driving source of turbulence. 

The 1D steady-state transport equation for the variance of density fluctuations can be written as \citep{2017ApJ...835..147Z,2018ApJ...854...32Z,2017ApJ...841...85A}
\begin{equation}
\begin{split}
U \frac{d}{dr} \langle {\rho^\infty}^2 \rangle + \frac{4U}{r} \langle {\rho^\infty}^2 \rangle + \frac{4}{r} \langle {u^\infty}^2 \rangle^{1/2} \langle {\rho^\infty}^2 \rangle = - \alpha \frac{\langle {u^\infty}^2 \rangle^{1/2} \langle {\rho^\infty}^2 \rangle}{\lambda^\infty_u} + \eta_1 \langle {\rho^\infty}^2 \rangle_0 \frac{r_0^2 |\Delta U|}{r^3},
\end{split}
\end{equation}
where $\eta_1$ is constant, and $\langle {\rho^\infty}^2 \rangle_0$ is the variance of density fluctuations at $r_0$. The second term on the right hand side is a shear driving source of turbulence for the density variance. The quasi-2D fluctuating velocity variance $\langle {u^\infty}^2 \rangle$, and the correlation length $\lambda_u^\infty$ can be expressed as,
\begin{equation}
\langle {u^\infty}^2 \rangle = \frac{\langle {{z^\infty}^+}^2 \rangle + \langle {{z^\infty}^-}^2 \rangle + 2E_D^\infty}{4}\ \text{and}\ \lambda_u^\infty = \frac{ (E_T^\infty+E_C^\infty) \lambda^+_\perp + (E_T^\infty - E_C^\infty) \lambda^-_\perp + E_D^\infty \lambda_D^\infty}{2(E_T^\infty + E_D^\infty)},
\end{equation}   
where $\lambda_\infty^+ (\equiv L_\infty^+/{\langle {{z^\infty}^+}^2 \rangle})$, $\lambda_\infty^-(\equiv L_\infty^-/{\langle {{z^\infty}^-}^2 \rangle})$ and $\lambda_D^\infty(\equiv L_D^\infty/E_D^\infty)$ are the quasi-2D correlation lengths corresponding to forward and backward propagating modes, and the residual energy, respectively.
 
The 1D steady-state transport equation for the solar wind proton temperature is given by
\begin{equation}
\begin{split}
U \frac{dT}{dr}  +(\gamma-1) \frac{2 U T}{r}   = \frac{s_1}{3} \frac{m_p}{ k_B} {\alpha} \bigg[ \frac{2 \langle {{z^*}^+}^2 \rangle {\langle {{z^\infty}^-}^2 \rangle}^{1/2}}{\lambda^+_\infty} + \frac{2 \langle {{z^*}^-}^2 \rangle {\langle {{z^\infty}^+}^2 \rangle}^{1/2}}{\lambda^-_\infty} \\
+ E^*_D \bigg (\frac{{ \langle {{z^\infty}^-}^2 \rangle}^{1/2} } {\lambda^+_\infty} +  \frac{ {\langle {{z^\infty}^+}^2 \rangle}^{1/2} }{\lambda^-_\infty}  \bigg )  + \frac{2 \langle {{z^\infty}^+}^2 \rangle {\langle {{z^\infty}^-}^2 \rangle}^{1/2}}{\lambda^+_\infty}  \\
+ \frac{2 \langle {{z^\infty}^-}^2 \rangle {\langle {{z^\infty}^+}^2 \rangle}^{1/2}}{\lambda^-_\infty} + E^\infty_D \bigg (\frac{{ \langle {{z^\infty}^-}^2 \rangle}^{1/2} }{\lambda^+_\infty} + \frac{ {\langle {{z^\infty}^+}^2 \rangle}^{1/2} }{\lambda^-_\infty}  \bigg )   \bigg],
\end{split}
\end{equation} 
where $m_p$ is the proton mass, $k_B$ is the Boltzmann constant, and $\gamma=5/3$ is a polytropic index. Here we assume that some fraction of the dissipated turbulent energy heats the solar wind plasma by introducing the parameter $s_1 (<1$). It is likely that some fraction of the turbulence energy goes into electron heating, some into creating a nonthermal population of ions (e.g., stochastic acceleration by magnetic islands), and some into creating a nonthermal electron population as well as of course heating ions. A posteriori, we find that if $s_1=0.3-0.4$, the heating prediction will be consistent with the temperature observed by PSP. Thus, we can conclude that $\sim 30-40$ \% of the available turbulent energy is used to heat the solar wind protons/ions. 

The turbulence transport model equations derived from the NI MHD equations contains several parametrizations and the Kolmogorov phenomenology is implicit in the model. One limitation is that the turbulence transport model equations assume certain symmetries of the turbulence in order to affect closure of the \textquotedblleft moments\textquotedblright~analysis of the transport equations. A second limitation is our use of a Kolmogorov phenomenology to model the nonlinear triple-correlation dissipation terms. In some cases, it may be more appropriate to utilize an Iroshnikov-Kraichnan-like phenomenology or a combination of Kolmogorov and Iroshnikov-Kraichnan phenomenology, which would introduce the Alfv\'en speed into the dissipation terms. Alternatively, some form of anisotropic dissipative cascade model might be considered \citep{2006PhRvL..96k5002B}. A third limitation is that the turbulence source terms are entirely phenomenological. The stream-shear source of turbulence in \cite{2017ApJ...835..147Z} is derived from dimensional analysis, for example.

Conversely, the turbulence transport theory presents several advantages. Firstly, the derived moment description captures the whole dynamic range of turbulence evolution in the inhomogeneous solar wind (i.e, coupled evolution of inwardly and outwardly propagating energy densities, coupled evolution of kinetic and magnetic energy densities, the coupled evolution of relevant correlation lengths, and all derived quantities) and the dynamic evolution of the dissipation rate in an inhomogeneous plasma with a turbulence source. Secondly, since the nonlinear term is obtained by adopting Kolmogorov phenomenology, it is not necessary to understand the detailed microphysics of dissipation and heating. This yields a robust determination of the solar wind heating rate. In principle, more detailed turbulence phenomenologies are readily incorporated. Finally, the nearly incompressible framework yields the evolution of weakly compressible quantities, such as the variance of the density fluctuations.  

\section{Results: Comparison with PSP-SWEAP plasma data}
In this section, we compare the theoretical solutions for the fluctuating kinetic energy, the correlation length of velocity fluctuations, the variance of density fluctuations, and the solar wind proton temperature with the PSP SWEAP plasma data, between $\sim 35.55$ R$_\odot$ and $\sim 131.64$ R$_\odot$. We solve the coupled turbulence transport model equations (4) --(14) using a Runge Kutta 4th order method. In the next section, we provide solutions of the turbulence transport model equations that include the fluctuating magnetic field. As noted, the turbulence transport solutions are best expressed in terms of the Els\"asser variables and we solve for the energy in forward and backward propagating modes, the total turbulent energy, normalized residual energy and the normalized cross-helicity, fluctuating magnetic energy, and correlation lengths corresponding to forward and backward propagating modes, residual energy and fluctuating magnetic energy. The coupled turbulence transport model equations are solved using the boundary conditions at $\sim 35.55$ R$_\odot$ shown in Table 2. To derive the boundary conditions for the majority quasi-2D and minority slab Els\"asser energies and residual energy, we assume an 80:20 ratio between the quasi-2D and slab turbulence. This choice is motivated by the original theoretical results of \cite{1992JGR....9717189Z}, who predicted that the ratio between quasi-2D and slab turbulence in fully developed slow wind turbulence is 80:20. Later, \cite{1996JGR...101.2511B} confirmed this ratio observationally. Observational studies \citep{2007ApJ...654L.103O,2009JGRA..114.7213W} also show that the slab correlation scale is about twice as large as the 2D correlation scale. Accordingly, we assume the ratio between the quasi-2D and slab correlation function is 2:1 to obtain the boundary conditions for the correlation function. The ratio between the quasi-2D and slab turbulence energy may change and will therefore result in different radial profiles for the quasi-2D and slab energy. However, the total turbulent energy remains the same. In the fast wind, for which turbulence is not fully developed, slab turbulence is the dominant component rather than the quasi-2D component. Table 3 shows the parameters used in the coupled turbulence transport model equations.
\begin{table}[h!]
	\centering
	\begin{tabular}{c c  | c c }
		\hline
		\multicolumn{2}{l|}{2D Core Model Equations}&\multicolumn{2}{l}{Slab Model Equations}\\
		\hline
		$\langle{{z^\infty}^+}^2\rangle$ & 9338.4 km$^2$s$^{-2}$ & $\langle{{z^*}^+}^2\rangle$ & 2334.6 km$^2$s$^{-2}$ \\
		$\langle{{z^\infty}^-}^2\rangle$ & 952.4 km$^2$s$^{-2}$  & $\langle{{z^*}^-}^2\rangle$ & 238.1 km$^2$s$^{-2}$ \\
		$E_D^\infty$ & -112.48 km$^2$s$^{-2}$ & $E_D^*$ & -28.12 km$^2$s$^{-2}$ \\
		$L_\infty^+$ & 5.19 $\times 10^8$ km$^3$s$^{-2}$ & $L_*^+$ & 2.59 $\times 10^8$ km$^3$s$^{-2}$\\
		$L_\infty^-$ & 5.44 $\times 10^7$ km$^3$s$^{-2}$ & $L_*^+$ & 2.72 $\times 10^7$ km$^3$s$^{-2}$ \\
		$L_D^\infty$ & -1.34 $\times 10^8$ km$^3$s$^{-2}$ & $L_D^*$ & -6.7 $\times 10^7$ km$^3$s$^{-2}$ \\
		$\langle {\rho^\infty}^2 \rangle$ & $2.83 \times 10^3$ cm$^{-6}$ &  &  \\
		$T$ & $1.75 \times 10^5 $ K &  &  \\
		\hline
	\end{tabular}
	\caption{Boundary values at 0.165 au ($35.55$ R$_\odot$) as measured by PSP at its closest approach to the Sun.}
\end{table}

\begin{table}[h!]
	\centering
	\begin{tabular}{c  c | c  c }
		\hline
		\multicolumn{2}{l|}{Parameters \quad Values} &\multicolumn{2}{l}{Parameters \quad Values}\\		\hline
		$C_{sh}^+$ & \quad  0.25  & 	$C_{sh}^{+*}$ & 0.2 \\
		$C_{sh}^-$ & \quad 0.1 & $C_{sh}^{-*}$ & 0.05 \\
		$C_{sh}^{E_D}$ & \quad -0.006 & $C_{sh}^{E_D*}$  & -0.003  \\
		U & \quad 380.0 km s$^{-1}$ & $\eta_1$ & 0.8\\
		$\Delta U$ & \quad 200.0 km s$^{-1}$ & $b$ & 0.26 \\
		$V_{A0}$ & \quad 101.37 km s$^{-1}$ & $\alpha $ & 0.1 \\
		$r_0$ & \quad 0.165 au & $n_{sw}$ & 232.34 \\ 
		\hline
	\end{tabular}
	\caption{Model parameters. }
\end{table}
The observed T- and N-component slow solar wind speed and solar wind density are shown in light blue in Figure 1 and were selected for comparison to the turbulence transport model described above. Here we selected those intervals that do not have any data gaps and contain the same resolution. Figure 1 shows the solar wind speed (top panel), the solar wind density (second panel), the solar wind proton temperature (third panel), and the mass flux (fourth panel), and the thermal plasma beta (fifth panel) measured by PSP as a function of heliocentric distance \citep{Casper2019}. These plasma parameters are moment data derived from the PSP SWEAP measurements. In the figure, the light blue corresponds to solar wind speed below 420 kms$^{-1}$. As shown in Figure 1, top panel, the observed slow radial velocity as well as T- and N-component solar wind speed ($U_T$ and $U_N$) appears to separate quite clearly into two components, one with a speed of $\sim 400$ kms$^{-1}$ and the second with speed $\sim 600$ kms$^{-1}$. As illustrated in the panels below, the remaining fluid variables tend to reflect these two classes of slow and fast speed solar wind. In the slow solar wind, turbulence is thought to be fully developed, in part due to its possible origin \citep[e.g.,][]{2003JGRA..108.1157F}. By contrast the fast solar wind tends to have outwardly propagating Alfv\'en waves only \citep[e.g.,][]{2005LRSP....2....4B,zhao2019}, again due to the nature of the origin of the fast wind \citep{2018ApJ...854...32Z,2019ApJ...876...26A,adhikari2019}. The analysis presented here considers only the evolution of turbulence in the slow solar wind, and we use the innermost PSP measurements at 35.55 R$_\odot$ as our boundary conditions. This choice is driven by the fact that most of the data from the first orbit is slow wind, and that we will have to wait for more data before we can start looking at other conditions (e.g., fast wind) with any statistical confidence. The dark blue in Figure 1 corresponds to the solar wind speed above 420 kms$^{-1}$. Compared with the slow solar wind, the density of the fast solar wind is smaller and the proton temperature a little higher. The top and second panels show the inverse relationship between the solar wind speed and the solar wind density. In the figure, the vertical line indicates the position of the CME observed by the PSP at $\sim 55.2$ R$_\odot$ \citep{Giacalone2019}. We exclude the CME in our calculation. 

Figure 1 also shows that the data can display a sharp changes, which may be a sign of boundary crossings of structures, such as pressure-balanced structures (PBSs) or flux tubes \citep{1968SoPh....4...67B,1995ISAA....3.....B,Vellante,1991JGR....96.1737B,2008JGRA..113.8110B,2014ApJ...783...65S}. The pressure-balanced structure is an equilibrium solution of NI MHD \citep{1992JGR....9717189Z}. A flux tube can be defined by a pressure-balanced structure since the pressure represents a smooth surface that is everywhere tangent to the local magnetic field \citep[see Appendix of][]{2004JGRA..109.4107Z}. In a companion PSP paper \cite{zhao2019} have identified numerous flux ropes (quasi-2D structures) in the slow wind observed by PSP over a wide range of scales, indicating the presence of quasi-2D turbulence. Thus, PBSs/flux tubes are part of the NI MHD description but, unlike the (unrealistic) static model of flux tubes discussed by \cite{2008JGRA..113.8110B} and others, PBSs/flux tubes are highly dynamical in the presence of quasi-2D turbulence \citep[Appendix A,][]{2004JGRA..109.4107Z}. The turbulence transport theory, which computes energy densities that are derived from NI MHD by taking moments, includes the dynamical role of these structures. These structures interact dynamically on a nonlinear time scale.   
\begin{figure}[h!]
	\centering
	\includegraphics[height=0.9\textwidth]{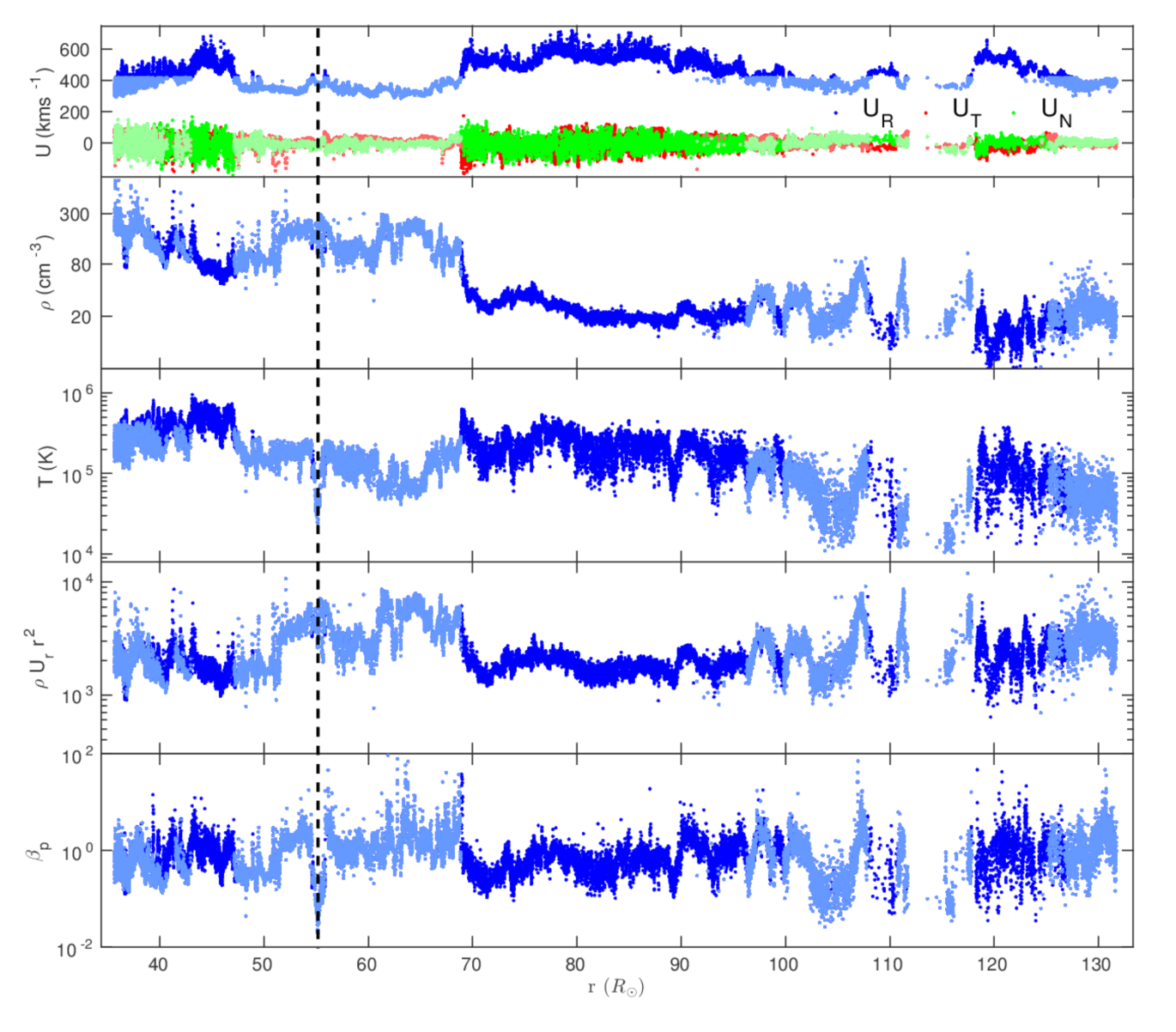}
	\caption{Parker Solar Probe SWEAP measurements: Solar wind speed (top panel), solar wind density (second panel), solar wind proton temperature (third panel), mass flux (fourth panel), and the plasma beta (bottom panel) as a function of heliocentric distance. Light blue \textquotedblleft.\textquotedblright~symbols correspond to solar wind speed less than $ 420$ kms$^{-1}$, and dark blue \textquotedblleft.\textquotedblright~symbols correspond to flow speeds greater than $420$ kms$^{-1}$. The vertical dashed line indicates a position of CME observed by PSP at $\sim 55.2$ R$_\odot$.}
\end{figure}

Following a similar procedure as in our previous papers \citep[e.g.,][]{2014ApJ...793...52A,2015ApJ...805...63A,2017ApJ...841...85A,2018ApJ...856...94Z}, we calculate the fluctuating kinetic energy, the correlation length of velocity fluctuations and the variance of the density fluctuations using a four hours moving interval, and then we smooth the observed quantities.
Figure 2 shows a comparison between the theoretical and observed fluctuating kinetic energy (left panel) and the correlation length of velocity fluctuations (right panel) as a function of heliocentric distance for those intervals corresponding to our identified slow wind intervals of Figure 1. The observed fluctuating kinetic energy and the corresponding correlation are shown with error bars. Here, the error bar denotes the standard deviation, i.e., the deviation of the data from the mean value. The error bars corresponding to the correlation lengths are larger than those of the fluctuating kinetic energy. In the figure, the dashed lines denote the fluctuating slab kinetic energy and the corresponding correlation length, the solid lines the fluctuating quasi-2D kinetic energy and the corresponding correlation length, and the dashed-dotted-dashed curve the total (quasi-2D plus slab) fluctuating kinetic energy. The theoretical fluctuating quasi-2D, slab, and total kinetic energy decrease approximately as $r^{-1.62}$, $r^{-1.18}$, and $r^{-1.47}$ with increasing heliocentric distance. The observed fluctuating kinetic energy also shows a decreasing profile with distance.
\begin{figure}[h!]
	\centering
	\includegraphics[height=0.36\textwidth]{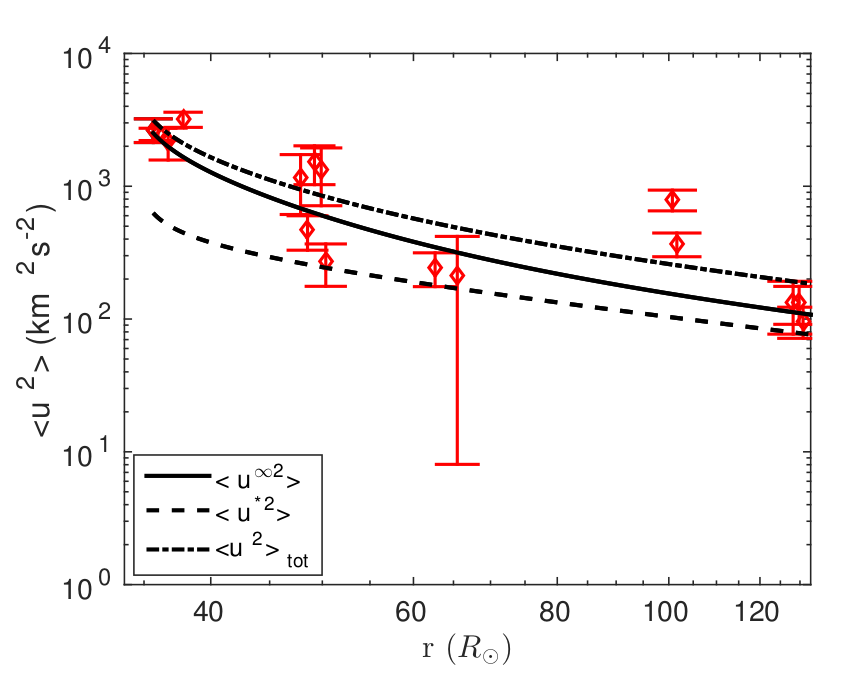}
	\includegraphics[height=0.35\textwidth]{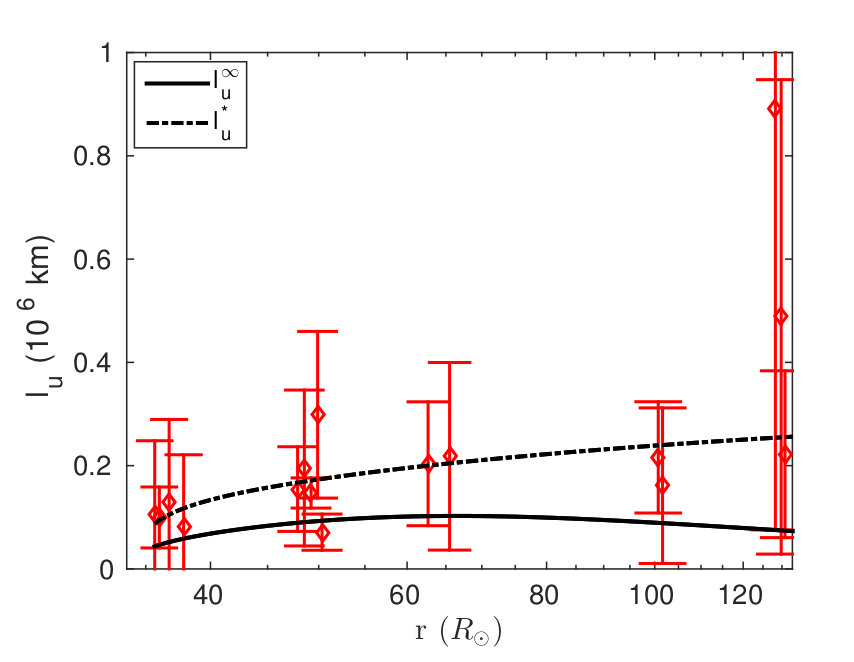}
	\caption{Comparison between the theoretical and observed fluctuating kinetic energy (left) and the corresponding correlation length (right) as a function of heliocentric distance. Solid curves represent the quasi-2D component, dashed curves the slab component, and dashed-dotted-dashed curve denote the total (quasi-2D plus slab) component. The red \textquotedblleft diamond\textquotedblright~symbols are observed results with error bars. }
\end{figure}

In the right panel of Figure 2, the theoretical correlation length for the quasi-2D fluctuating kinetic energy (solid curve) increases until $\sim 65$ R$_\odot$, and then decreases slightly as distance increases. The correlation length corresponding to the slab fluctuating kinetic energy (dashed curve) increases with heliocentric distance. The theoretical correlation length corresponding to the slab fluctuating kinetic energy is larger than that of the quasi-2D fluctuating kinetic energy between the heliocentric distance of $\sim 35.55$ R$_\odot$ and $\sim 131.64$ R$_\odot$. Similarly, the fluctuating quasi-2D kinetic energy is larger than the fluctuating slab kinetic energy in the same region. As discussed above, this is due to the assumed boundary conditions between the quasi-2D and slab
turbulence at 35.55 R$_\odot$. We assume an 80:20 ratio between the quasi-2D and slab turbulence energy \citep{1992JGR....9717189Z,1996JGR...101.2511B} and a 2:1 ratio between the correlation lengths of slab and quasi-2D turbulence \citep{2007ApJ...654L.103O,2009JGRA..114.7213W}, which is a well established hypothesis for a fully developed slow wind turbulence. However, this ratio may change depending on different conditions, such as the solar cycle. The theoretical fluctuating kinetic energy and the correlation length as a function of heliocentric distance are obtained by using Equation (13), which requires the Els\"asser energies, residual energy, and the corresponding correlation functions. These quantities are obtained by solving coupled quasi-2D and slab turbulence transport equations (4)--(11) using boundary conditions shown in Table 2. Table 3 lists the theoretical values and the observed values with error at the boundary (35.55 R$_\odot$), and shows that the theoretical values of the fluctuating quasi-2D, slab, and total kinetic energy at 35.55 R$_\odot$ are within the error bar of the observed fluctuating kinetic energy. The error bar in the correlation length plot is very large. Although, the error in the correlation length is very large, the theoretical quasi-2D and slab correlation length of velocity fluctuations at the left boundary is within the error bar of the observed correlation. This suggests that the boundary conditions shown in Table 2 are close to those of the observed quantities.    
Our results would indicate that quasi-2D turbulence is dominant rather than slab turbulence in a fully developed slow wind turbulence between 35.55 R$_\odot$ and 131.64 R$_\odot$ \citep{1992JGR....9717189Z,2017ApJ...835..147Z,2018ApJ...856...94Z,1996JGR...101.2511B,2017ApJ...841...85A}. In the fast wind, slab turbulence (or even uni-directionally propagating Alfv\'en waves) dominates quasi-2D turbulence in the inner heliosphere \citep[e.g.,][]{2005ApJ...635L.181D}.

\begin{figure}[h!]
	\centering
	\includegraphics[height=0.36\textwidth]{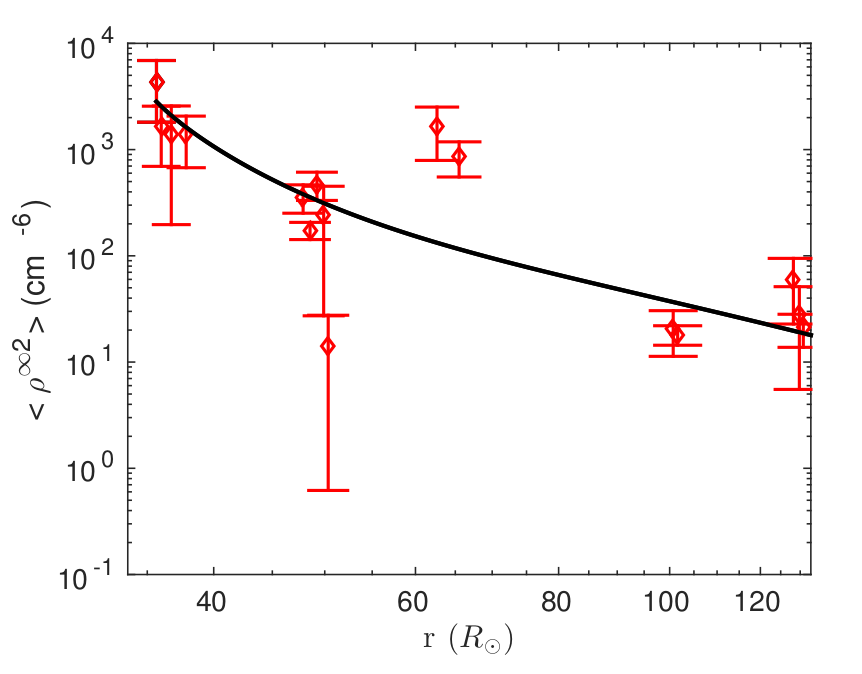}
	\includegraphics[height=0.36\textwidth]{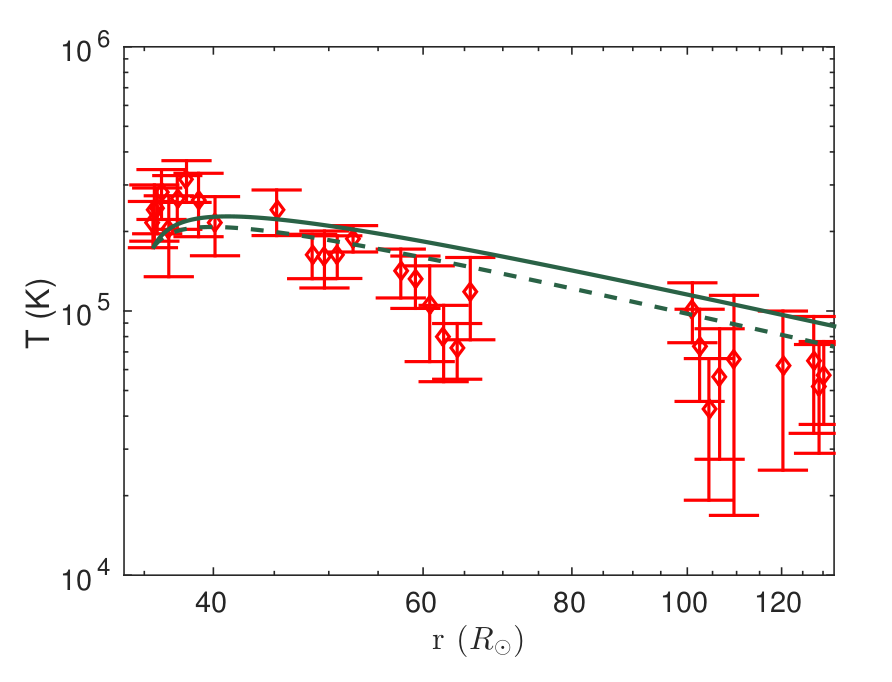}
	\caption{Left: Comparison of the theoretical density variance and observed density variance as a function of heliocentric distance. Right: Comparison of the theoretical and observed solar wind proton temperature with heliocentric distance. The solid and dashed green curves in the right panel correspond to $s_1=0.4$ and $s_1=0.3$, respectively. The red \textquotedblleft diamond\textquotedblright~symbols are observed results with error bars. }
\end{figure}

\begin{table}[h!]
	\begin{center}
		\label{tab:table1}
		\begin{tabular}{l c r} 
			\hline
			\text{Parameters } & \text{Theoretical values} & \text{Observed values $\pm$ $\sigma$}\\
			\hline
			$\langle u^2 \rangle_{tot}$ (km$^2$s$^{-2}$) & $3.15 \times 10^3$ & $2.68\times10^3 \pm 542.1$  \\
			$\langle {u^\infty}^2 \rangle$ (km$^2$s$^{-2}$) & $2.52 \times 10^3$ & $2.46\times10^3 \pm 433.68$ \\
			$\langle {u^*}^2 \rangle$ (km$^2$s$^{-2}$) & $629$ &  $536 \pm 108.42 $  \\
			$l_u^\infty$ (km) & $0.0436 \times 10^6$ & $0.11 \times 10^6 \pm 0.14 \times 10^6$ \\
			$l_u^*$ (km) & $0.0872 \times 10^6$ &  $0.11 \times 10^6 \pm 0.14 \times 10^6$ \\
			$\langle {\rho^\infty}^2 \rangle $ (cm$^{-6}$) & $4.35 \times 10^3$ &  $4.34 \times 10^3 \pm 2.5 \times 10^3$\\
			$T$ (K) & $1.75 \times 10^5$ &  $2.17 \times 10^5 \pm 4.31 \times 10^4$\\
			\hline
		\end{tabular}
		\caption{Theoretical values and observed values with errors at 35.55 R$_\odot$. The parameter $\sigma$ denotes the standard deviation. }
	\end{center}		
\end{table}

The left panel of Figure 3 shows a comparison between the theoretical and observed variances of density fluctuations as a function of heliocentric distance. In the figure, the theoretical variance of density fluctuations (solid curve) decreases monotonically in a manner similar to that of observed density variance (red \textquotedblleft diamond\textquotedblright with error bar) with increasing heliocentric distance. The theoretical density variance exhibits a radial profile of $r^{-2.98}$. The rate at which the variance of density fluctuations decreases in the inner heliosphere is slower than that of the outer heliosphere \citep{2017ApJ...835..147Z,2018ApJ...856...94Z,2017ApJ...841...85A}, which decreases faster than $r^{-3}$. Table 4 shows that the theoretical variance of density fluctuations at 35.55 R$_\odot$ is within the error bar of the observed density variance.

The right panel of Figure 3 displays the solar wind proton temperature. The observed solar wind proton temperature with error is calculated for $\sim 5.83$ hours intervals. The comparison between the theoretical and observed solar wind proton temperature shows that the theoretical solar wind proton temperature is a little larger than that of observed solar wind proton temperature. In the figure, the solid green curve corresponds to the $s_1=0.4$ and the dashed green curve to $s_1=0.3$. It indicates that, in the former case, 40 \% of the turbulence energy goes into solar wind heating of the protons, while in the latter case, 30 \% of the turbulence heating goes into solar wind heating of the protons. This result would suggest that the remaining energy in turbulence fluctuations may be dissipated into electron heating and the non-thermal energization of charged particles. The solid and dashed green curves increase initially and then decrease as $r^{-0.89}$ and $r^{-0.95}$, respectively, indicating that the heating rate determines the radial profile of the solar wind proton temperature. The rate of cooling for the theoretical solar wind proton temperature is slower than that of adiabatic cooling indicating that the energy is being added in situ, either through continued dissipation of turbulence, or the generation of in situ turbulence and its subsequent dissipation. The cooling rate depends on the cascade rate \citep{2010JGRA..115.2101N}, which is, in this manuscript, based on Kolmogorov phenomenology, but other phenomenologies may also be applied, such as Iroshnikov-Kriachnan, for example \citep{2010JGRA..115.2101N}. Here, the chosen boundary value of the solar wind proton temperature is within the error bar of the observed solar wind proton temperature as shown in Table 4. 
\section{Results: Theoretical predictions}
Having extracted the plasma variables from the solutions of the equations describing the Els\"asser variables (4)--(12), and compared them to PSP SWEAP observations, we can predict the corresponding turbulent Els\"asser and magnetic variables. Figure 4 shows the theoretical turbulent quantities as a function of heliocentric distance. The solid curves denote the majority quasi-2D components, the dashed curves the minority slab component, 
\begin{figure}[h!]
	\centering
	\includegraphics[height=0.28\textwidth]{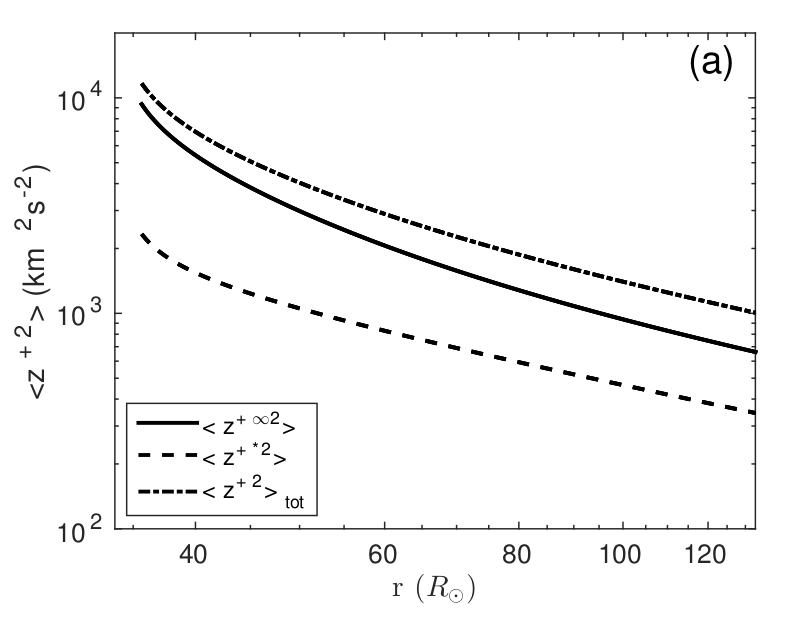}
	\includegraphics[height=0.28\textwidth]{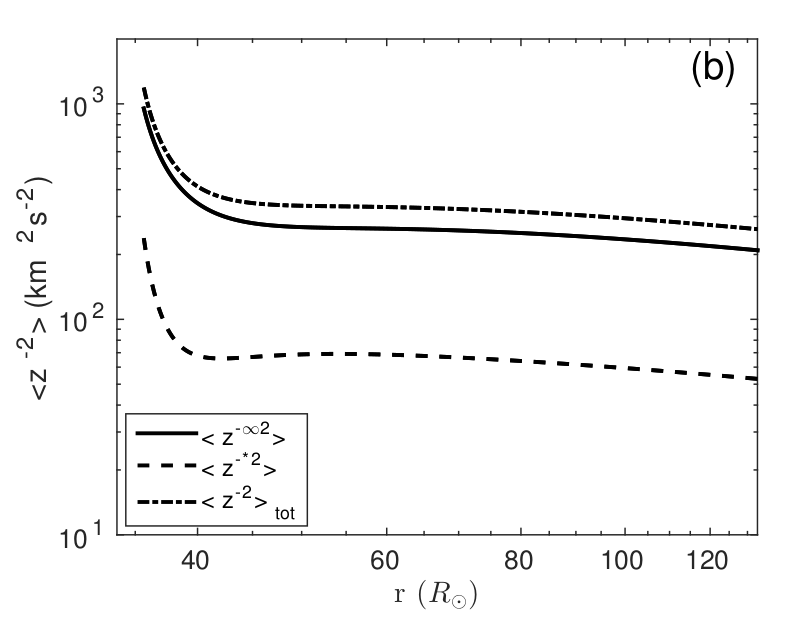}
	\includegraphics[height=0.28\textwidth]{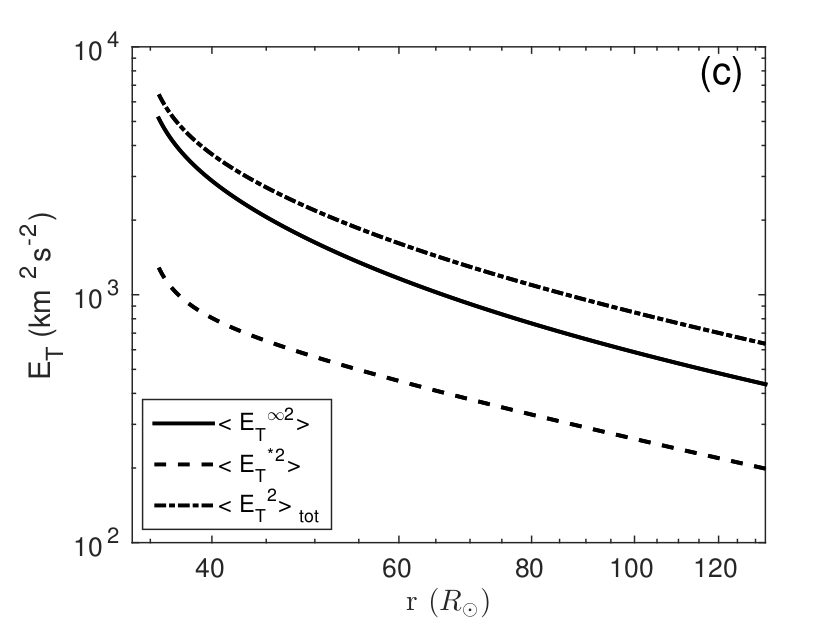}
	\includegraphics[height=0.28\textwidth]{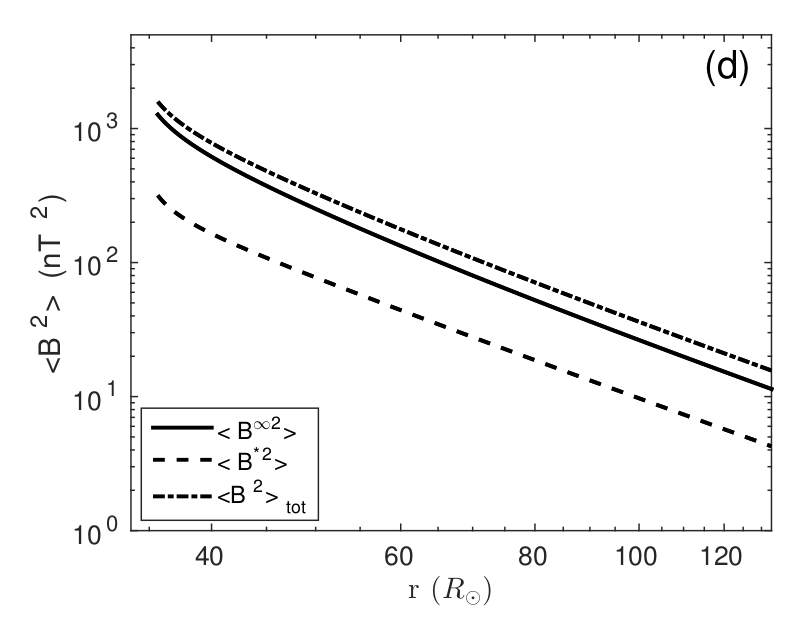}
	\includegraphics[height=0.28\textwidth]{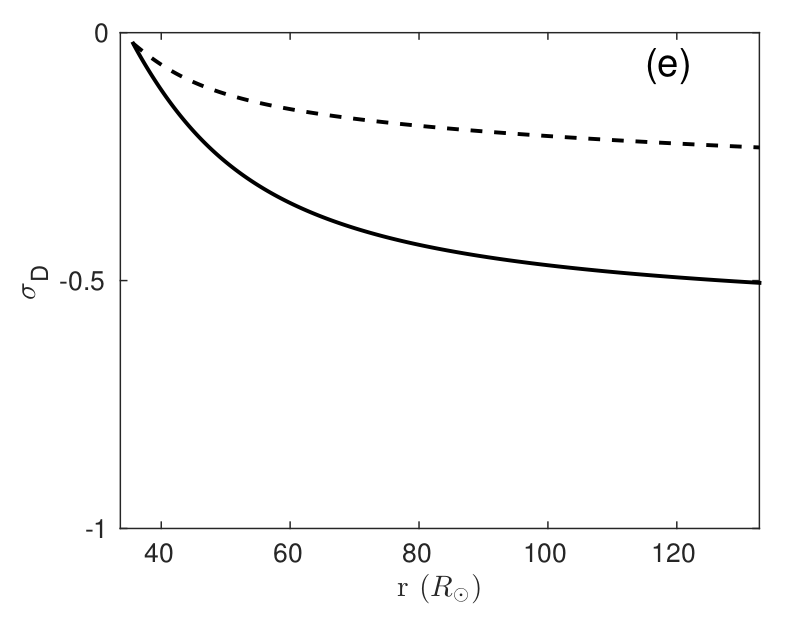}
	\includegraphics[height=0.28\textwidth]{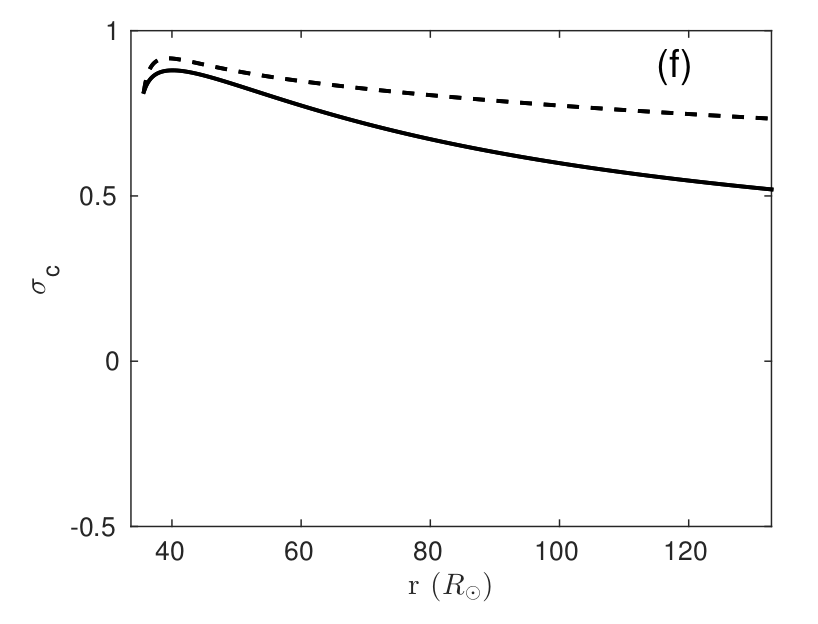}
	\includegraphics[height=0.28\textwidth]{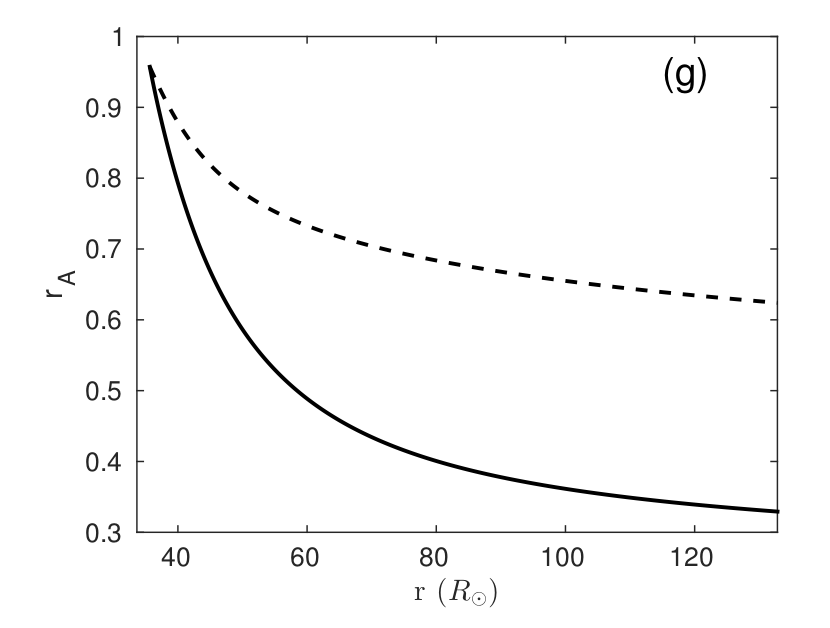}
\caption{Turbulent quantities as a function of heliocentric distance. The convention for the curve is the same as used in Figure 2. The panels show (a) the energy in forward propagating modes, (b) the energy in backward propagating modes, (c) the total turbulent energy, (d) the fluctuating magnetic energy, (e) the normalized residual energy, (f) the normalized cross-helicity, and (g) the Alfv\'en ratio. }
\end{figure}
and the dashed-dotted-dashed curves the total (sum of quasi-2D and slab) component. Figure 4a shows the decay of the turbulent energy in forward propagating modes, with power law decreases as $r^{-1.45}$, $r^{-1.13}$, and $r^{-1.36}$, corresponding to quasi-2D, slab, and total components, respectively. Figure 4b shows the energy in backward propagating modes as a function of heliocentric distance. In this case, the quasi-2D, slab, and total energy in backward propagating modes decrease rapidly initially, increase slightly, and then decrease slightly as $r^{-0.39}$, $r^{-0.39}$, and $r^{-0.4}$. The slight increase in backward propagating modes is due to the excitation of these modes by stream shear \citep{2015ApJ...805...63A}. Figure 4c shows the total turbulent energy as a function of heliocentric distance. Similar to the energy in forward propagating modes, the total turbulent energy decreases monotonically with heliocentric distance. Here, the quasi-2D, slab and (quasi-2D + slab) total turbulent energies follow radial profiles of $r^{-1.27}$, $r^{1.1}$, and $r^{-1.21}$, respectively. 

Figure 4d shows that the fluctuating magnetic energy decreases with the increase of heliocentric distance, as (quasi-2D, slab, and total) $r^{-3.12}$, $r^{-2.97}$, and $r^{-3.1}$ respectively, indicating that the radial profile of the fluctuating magnetic energy is approximately similar to that of the well-known WKB description \citep{1996JGR...10117093Z}. The latter result is rather interesting since an $r^{-3}$ decay corresponds to the variance of the fluctuating magnetic field is what is expected of WKB theory \citep{1996JGR...10117093Z}. However, WKB is a linear Alfv\'en wave theory (i.e., non-interacting Alfv\'en modes and therefore does not describe turbulence) and the prior results (observations + theory) are certainly inconsistent with WKB theory. For example, Figure 2 shows that $\langle u^2 \rangle$ decays as $r^{-1.47}$, whereas of WKB theory, since $\langle u^2 \rangle = \langle b^2 \rangle$, should decay as $r^{-3}$. As discussed and illustrated in Zank et al. (1996), an $r^{-3}$ decay is the fluctuating magnetic field variance emerges naturally from a turbulence transport formalism when the rate of dissipation is balanced by the rate of turbulence driving-- in this case the turbulence is driven by shear on the boundaries of the fast and slow streams. The difference between the decay characteristics of $\langle u^2 \rangle$ and $\langle b^2 \rangle$ is interesting in that it shows that it is primarily magnetic energy rather than kinetic energy that dominates at this distance. 

Figure 4e shows the normalized residual energy as a function of heliocentric distance. The normalized residual energy for quasi-2D and slab turbulence decreases with increasing heliocentric distance, as does the normalized cross-helicity (Figure 4f). The evolution of the normalized cross-helicity shows that the slab turbulence remains essentially outwardly propagating Alfv\'en modes, and there is relatively little generation of inwardly propagating Alfv\'en waves. By contrast, the quasi-2D turbulence evolution of the cross-helicity tends to smaller values more rapidly than the slab turbulence. The rapid decrease in the Alfv\'en ratio for the quasi-2D modes compared to that of the slab turbulence (Figure 4g) illustrates that the quasi-2D turbulence is more magnetically dominated than slab turbulence.  

\begin{figure}[h!]
	\centering
	\includegraphics[height=0.45\textwidth]{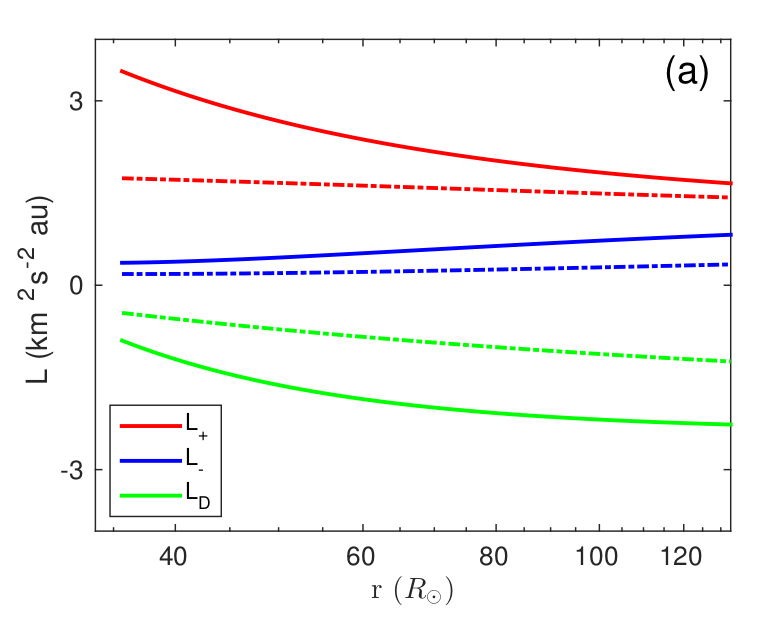}
	\includegraphics[height=0.38\textwidth]{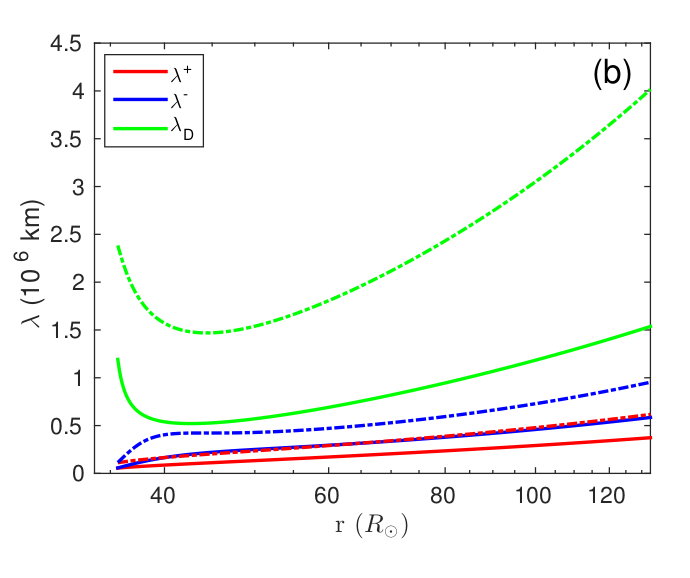}
	\includegraphics[height=0.38\textwidth]{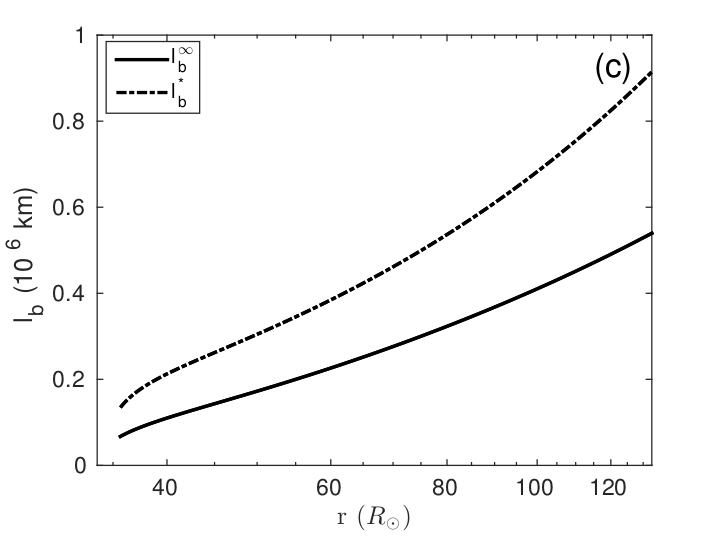}
	\caption{The panels a and b show the correlation functions and correlation lengths of forward and backward propagating modes, and the residual energy. The panel c shows the correlation length of magnetic field fluctuations. The description of curves is similar to Figure 2. }
\end{figure}
Figure 5a shows the correlation functions of the forward propagating modes (red curves), backward propagating modes (blue curves), and the residual energy (green curves) as a function of heliocentric distance. The solid curves identify the quasi-2D correlation functions, and the dashed curves the slab correlation functions. The quasi-2D correlation function for forward propagating modes decreases with heliocentric distance, and the slab correlation function decreases very slightly in the distance between $35.55$ R$_\odot$ and 131.64 R$_\odot$. Both the quasi-2D and slab correlation functions corresponding to backward propagating modes increase slightly with heliocentric distance. Similarly, the quasi-2D and slab correlation functions corresponding to the residual energy decrease as a function of heliocentric distance.

Figure 5b shows the correlation lengths corresponding to forward and backward propagating modes. The quasi-2D and slab correlation lengths for the forward propagating mode increase with distance as $r^{1.02}$ and $r^{0.98}$, respectively. Similarly, the quasi-2D and slab correlation lengths for backward propagating modes increase as $r^{0.94}$ and $r^{0.92}$, respectively. The quasi-2D and slab correlation lengths for the residual energy initially decrease, and then increase as $r^{0.93}$ and $r^{0.97}$, respectively.  
Figure 5c shows the correlation length for magnetic field fluctuations. The correlation length for quasi-2D magnetic field fluctuation shows a radial trend of $r^{1.12}$, while the correlation length for slab magnetic field fluctuations is $\sim r^{1.11}$, i.e., the basically exhibit the same radial dependence.
\section{Discussion and Conclusions}
We studied the evolution of turbulence in the inner heliosphere along the trajectory of the Parker Solar Probe (PSP) in the outbound direction from the perihelion ($\sim 35.55$ R$_\odot$) of the first orbit to the heliocentric distance of $\sim 131.64$ R$_\odot$, by using PSP SWEAP plasma measurements and a nearly incompressible magnetohydrodynamic Zank et al. (2017) turbulence transport model. Furthermore, based on the theory of Zank et al. (2017), we can predict additional turbulent quantities that include both the fluctuating velocity and magnetic field. For the present, we only compared the plasma quantities predicted by Zank et al. (2017) with the PSP SWEAP plasma measurements, and will compare the magnetic field data once it is available. Since the observed radial velocity and T- and N-component solar wind speed (and also the solar wind density) are seen to be divided clearly into two components, one with a speed of $\sim 400$ kms$^{-1}$ and other with a speed of $\sim 600$ kms$^{-1}$, we considered the solar wind speed, the density, and proton temperature corresponding to the slow solar wind regime, as determined by a speed less than 420 kms$^{-1}$. By doing so, we compared the theoretical and observed evolution of fully developed turbulence between $\sim 35.55$ R$_\odot$ and $\sim 131.64$ R$_\odot$.

We compared the theoretical and observed fluctuating kinetic energy, the correlation length of velocity fluctuations, the variance of density fluctuations, and the solar wind proton temperature. We found that the plasma quantities predicted by the model are in reasonable agreement with the observed PSP SWEAP plasma measurements. On the basis of these comparisons, other theoretical results related to the fluctuating magnetic field were derived, such as the energy in forward and backward propagating modes, the normalized residual energy and cross-helicity, the fluctuating magnetic energy, the total turbulent energy, the correlation functions corresponding to forward and backward propagating modes and the residual energy, the correlation length corresponding to forward and backward propagating modes and the residual energy, and the correlation length of magnetic field fluctuations between $\sim 35.55$ R$_\odot$ and $\sim 131.64$ R$_\odot$. In future work, we will compare the Els\"asser and magnetic field quantities predicted by our turbulence transport model with the corresponding quantities measured by the PSP SWEAP plasma and field measurements.    

We summarize our findings for the range between the perihelion of the first orbit of the PSP and $131.64$ R$_\odot$ as follows.
\begin{itemize}
	\item The theoretical and observed fluctuating kinetic energy decreases with increasing heliocentric distance. The theoretical quasi-2D, slab, and total fluctuating kinetic energy follow power laws of $r^{-1.65}$, $r^{-1.2}$, and $r^{-1.49}$, respectively.
	
	\item The correlation length for the theoretical and observed velocity fluctuations increases with increasing heliocentric distance.
	
    \item The theoretical and observed variance of the density fluctuations decreases with heliocentric distance. The theoretical variance of density fluctuations decreases as $r^{-2.91}$.	

    \item The theoretical and observed solar wind proton temperature decreases with distance, and we find that approximately 30--40 \% of the dissipated turbulent energy is sufficient to account for the observed proton temperature profile. That implies that approximately 70 \% of the turbulent energy is used to heat electrons and create energetic particle population.  
    
    \item The theoretical quasi-2D, slab and total turbulent energy in forward propagating modes predicts radial profiles of $r^{-1.48}$, $r^{-1.16}$, and $r^{-1.38}$, respectively.
    
    \item The quasi-2D, slab and total turbulent energy in backward propagating modes is predicted to decrease initially, increase slightly and then decrease as  $r^{-0.38}$, $r^{-0.38}$, and $r^{-0.38}$, respectively.
    
    \item The theoretical quasi-2D, slab and (quasi-2D + slab) total turbulent energy are predicted to decrease as $r^{-1.3}$, $r^{1.08}$, and $r^{-1.24}$, respectively.
    
     \item The quasi-2D, slab and total fluctuating magnetic energy are predicted to decay as power laws with the form $r^{-3.14}$, $r^{-2.99}$, and $r^{-3.1}$, respectively.
     
     \item The quasi-2D and slab correlation lengths corresponding to forward and backward propagating modes are predicted to increase with distance, whereas the correlation length for the residual energy is predicted to decrease initially, and then increase with distance .
     
     \item The correlation lengths corresponding to quasi-2D and slab fluctuating magnetic energy are predicted to increase according to $r^{1.15}$ and $r^{1.13}$, respectively.
     
     {\item The observed normalized density fluctuations $\delta \rho/\rho$ and the turbulent sonic Mach number ($M_s$) are small between 35.5 R$_\odot$ and 131.64 R$_\odot$. The scaling between the density fluctuations and the turbulent sonic Mach number is found to be $\delta \rho \sim O(M_s^{0.97})$ in this region, indicating that the nearly incompressible MHD theory is an appropriate model to describe turbulence in the solar wind.  }
\end{itemize}


\acknowledgments
\begin{acknowledgements}
	The SWEAP investigation and this publication are supported by the PSP mission under NASA contract NNN06AA01C. We acknowledge the partial support of NASA grants NNX 08AJ33G, Subaward
	37102-2, NNX14AC08G, the PSP contract SV4-84017, an NSF-DOE grant PHY-1707247, and the partial support of NSF EPSCoR RII-Track-1 cooperative agreement OIA-1655280. L.A. and L.L.Z. thank K.E. Korreck, A.W. Case and M. Stevens for their kind hospitality while visiting the Smithsonian Astrophysics Observatory (SAO).
\end{acknowledgements}

\newpage
\section*{APPENDIX A: Scaling of density fluctuations with turbulent sonic Mach number}
On MHD scales, the solar wind behaves as an almost incompressible fluid in both the inner and the outer heliopshere, with the relative amplitude of density fluctuations being less than 0.1 \citep{1995JGR...100.9475B}. Thus, nearly incompressible MHD theory seems to be applicable in describing much of solar wind turbulence at these scales. In this Appendix, we use PSP measurements in the region between 35.5 R$_\odot$ and 131.64 R$_\odot$ to determine whether the solar wind is compressible or nearly incompressible. The theory of NI MHD has been developed since the late 1980s \citep{1988PhFl...31.3634M,1991JGR....96.5421M,1991PhFl....3...69Z,1992JGR....9717189Z,1993PhFl....5..257Z,2006PhRvE..74b6302H,2010ApJ...718..148H,1998ApJ...494..409B}, and the theory predicts, i) that $\delta \rho$ scales as $\sim O(M_s^2)$ if the background flow is homogeneous \citep{1988PhFl...31.3634M,1991PhFl....3...69Z,1992JGR....9717189Z}, and ii) $\delta \rho$ scales as $\sim O(M_s)$ if the background field is inhomogeneous \citep{2006PhRvE..74b6302H,2010ApJ...718..148H,1998ApJ...494..409B}. Observational studies in the solar wind \citep{1993JGR....98.7837K,1995JGR...100.9475B,1994JGR....9921481T} find that the $O(M_s^2)$ scaling is met rarely and that an $O(M_s)$ scaling is more appropriate. This is consistent with NI MHD in an inhomogeneous flow.

To find a scaling between density fluctuations and the turbulent sonic Mach number, the density fluctuation and the turbulent sonic Mach number are calculated for four hour moving intervals in the slow solar wind plasma identified by light blue in Figure 1. The results are then smoothed by taking 20 data points. Figure 6 shows the relation between the density fluctuations ($\delta \rho$) and the turbulent sonic Mach number $M_s(=\delta u/C_s)$, where $\delta u$ is the characteristic speed of the fluctuations and $C_s=\sqrt{(\gamma P/\rho)}$ is the sound speed. Here $\gamma(=5/3)$ is the polytropic index, $P$ is the local solar wind thermal pressure, and $\rho$ is the local solar wind density. The black solid line is the least square fit of the $\delta \rho$ and $M_s$ scatter plot. We find that $\delta \rho \sim O(M_s^{0.97})$. This scaling is close to the $\delta \rho \sim O(M_s)$ scaling predicted by \cite{2006PhRvE..74b6302H,2010ApJ...718..148H,1998ApJ...494..409B} for an inhomogeneous background flow. 
\begin{figure}[h!]
	\centering
	\includegraphics[height=0.4\textwidth]{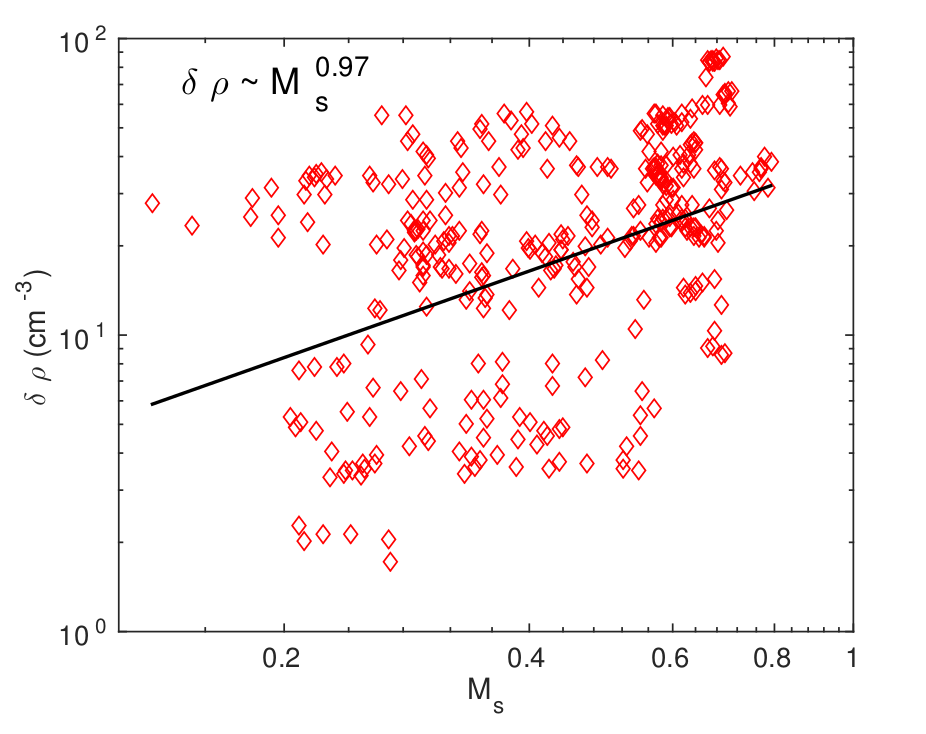}
	\caption{Density fluctuations $\delta \rho$ as a function of turbulent sonic Mach number $M_s$ for solar wind speeds less than 420 kms$^{-1}$. The black line is a least-square fit showing that $\delta \rho \sim M_s^{0.97}$.  }
\end{figure}

The frequency distributions of $\delta \rho/\rho$ and $M_s$ are shown in the left and right panel of Figure 7, respectively. The left panel of Figure 7 shows that $\delta \rho/\rho$ is concentrated mainly around $\sim 0.15$, which shows that the flow is essentially incompressible at the scale studied (4 hour intervals). The right panel of Figure 7 shows that the most likely value of the turbulent sonic Mach number distribution is bimodel, peaking near $\sim 0.3$ and $\sim 0.6$. These results suggest that the nearly incompressible approach is suitable for studying turbulence in the solar wind. The length of interval used may affect the result. One effect may be that the histogram of $\delta \rho/\rho$ over longer time intervals will move to the right \citep{1995JGR...100.9475B}.  
\begin{figure}[h!]
	\centering
	\includegraphics[height=0.35\textwidth]{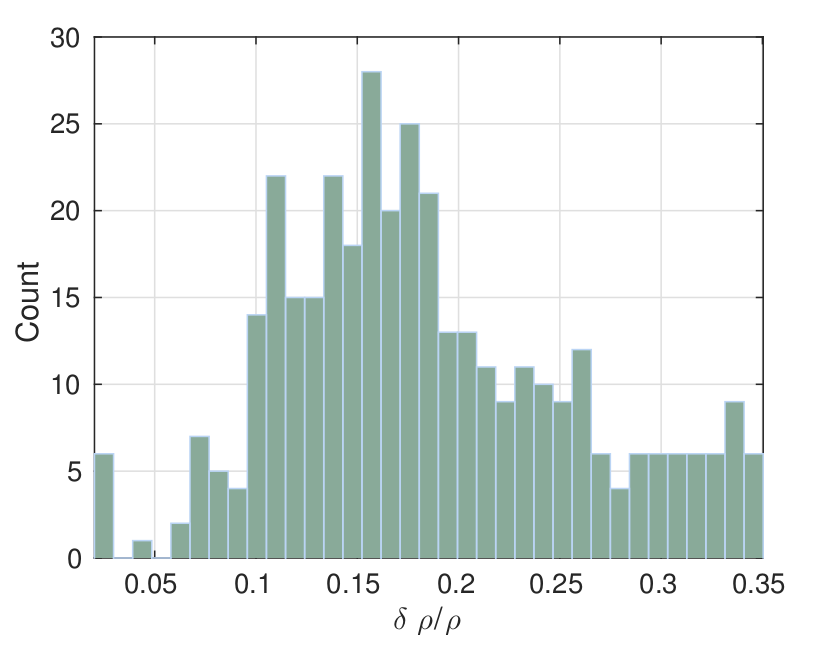}
	\includegraphics[height=0.35\textwidth]{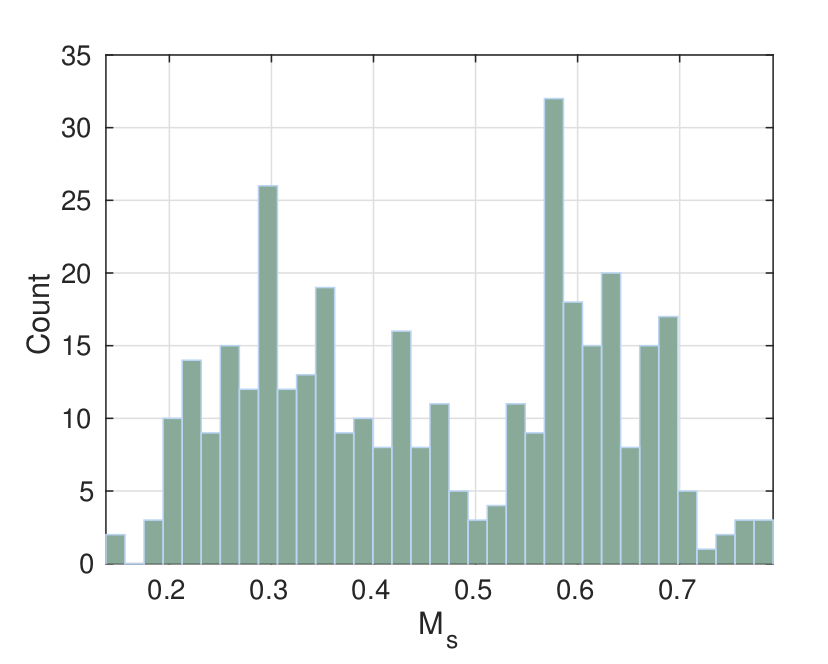}
	\caption{Left: Histogram of the density fluctuations normalized to the mean density. Right: Histogram of the turbulent sonic Mach number. }
\end{figure}


\end{document}